\documentclass[submission,copyright,creativecommons]{eptcs}
\providecommand{\event}{FMAS 2023} 

\usepackage{iftex}

\usepackage{amsmath,amssymb}       
\usepackage{cases}
\usepackage{empheq}
\usepackage{mathbbol}
\usepackage{bm}
\DeclareSymbolFontAlphabet{\amsmathbb}{AMSb}
\usepackage{mathtools}

\usepackage{tabularx}
\usepackage{booktabs}

\usepackage{caption}
\usepackage{subcaption}
\usepackage{appendix}
\usepackage{siunitx}

\usepackage{cleveref}

\crefname{equation}{Eq.}{Eqs.}
\crefname{pluralequation}{Eqs.}{Eqs.}

\crefname{algorithm}{Algorithm}{Algorithm}

\crefname{figure}{Fig.}{Figs.}
\crefname{pluralfigure}{Figs.}{Figs.}

\crefname{section}{Sect.}{Sects.}
\crefname{pluralsection}{Sects.}{Sects.}

\crefname{table}{Table}{Table}
\crefname{pluraltable}{Tables}{Tables}

\crefname{definition}{Def.}{Def.}
\crefname{pluraldefinition}{Defs.}{Defs.}

\crefname{theorem}{Theorem}{Theorems}
\crefname{pluraltheorem}{Theorems}{Theorems}

\crefname{lemma}{Lemma}{Lemmas}
\crefname{plurallemma}{Lemmas}{Lemmas}

\crefname{example}{Example}{Example}
\crefname{pluralexample}{Examples}{Examples}

\crefname{problem}{Problem}{Problem}
\crefname{pluralproblem}{Problems}{Problems}

\crefname{assumption}{Assumption}{Assumption}
\crefname{pluralassumption}{Assumptions}{Assumptions}

\crefname{remark}{Remark}{Remark}
\crefname{pluralremark}{Remarks}{Remarks}

\crefname{appendix}{Appendix}{Appendices}
\crefname{pluralappendix}{Appendices}{Appendices}

\usepackage{tikz} 
\usepackage{pgfplots}
\usepgfplotslibrary{external,fillbetween}
\pgfplotsset{compat=1.8}

\definecolor{red}{rgb}{0.745,0.192,0.102}
\definecolor{darkgreen}{RGB}{34,161,55}
\definecolor{ruhuisstijlrood}{rgb}{0.745,0.192,0.102}
\definecolor{ruhuisstijlzwart}{rgb}{0,0,0}
\definecolor{ruhuisstijlwit}{rgb}{0.98,0.98,0.98}
\definecolor{plotblue}{rgb}{0.1,0.498039215686275,0.9549019607843137}
\newcommand{\rured}[1]{\textcolor{ruhuisstijlrood}{#1}}
\newcommand{\darkgreen}[1]{\textcolor{darkgreen}{#1}}

\usepackage{mdframed}

\usetikzlibrary{patterns,arrows,arrows.meta,calc,shapes,shadows,decorations.pathmorphing,decorations.pathreplacing,automata,shapes.multipart,positioning,shapes.geometric,fit,circuits,trees,shapes.gates.logic.US,fit, shadows.blur,shapes.symbols}

\usepackage[acronym]{glossaries}

\usepackage{fontawesome5}
\newcommand{\ie}{i.e.\@}

\newcommand{\eg}{e.g.\@}
\newcommand{\prism}{\textrm{PRISM}}

\newcommand{\stochy}{\textrm{StocHy}}

\newcommand{\probreach}{\textrm{ProbReach}}

\newcommand{\R}{\amsmathbb{R}}

\newcommand{\N}{\amsmathbb{N}}

\newcommand{\controller}{\ensuremath{c}}

\newcommand{\cControlSpace}{\ensuremath{\mathcal{U}}}

\newcommand{\state}{\ensuremath{x}}
\newcommand{\control}{\ensuremath{u}}
\newcommand{\noise}{\ensuremath{\eta}}
\newcommand{\drift}{\ensuremath{q}}

\glsdisablehyper

\newacronym[]{LTL}{LTL}{linear temporal logic}
\newacronym[]{iid}{i.i.d.}{independent and identically distributed}
\newacronym[]{UAV}{UAV}{unmanned aerial vehicle}

\newacronym[]{PAC}{PAC}{probably approximately correct}
\newacronym[]{RL}{RL}{reinforcement learning}
\newacronym[]{DRO}{DRO}{distributionally robust optimization}
\newacronym[plural=MCs,firstplural=Markov chains (MCs)]{MC}{MC}{Markov chain}
\newacronym[plural=MDPs,firstplural=Markov decision processes (MDPs)]{MDP}{MDP}{Markov decision process}
\newacronym[plural=iMDPs,firstplural=interval Markov decision processes (iMDPs)]{iMDP}{iMDP}{interval Markov decision process}
\newacronym[plural=aMDPs,firstplural=augmented MDPs (aMDPs)]{aMDP}{aMDP}{augmented MDP}
\newacronym[plural=POMDPs,firstplural=partially observable Markov decision processes (POMDPs)]{POMDP}{POMDP}{partially observable Markov decision process}

\glsunset{iMDP}
\glsunset{iid}

\usepackage{microtype}

\ifpdf
  \usepackage{underscore}         
  \usepackage[T1]{fontenc}        
\else
  \usepackage{breakurl}           
\fi

\title{Correct-by-Construction Control for Stochastic and \\ Uncertain Dynamical Models via Formal Abstractions}

\author{Thom Badings \quad Nils Jansen
\institute{Radboud University\\Nijmegen, the Netherlands}
\email{thom.badings@ru.nl}
\and
Licio Romao \quad Alessandro Abate
\institute{University of Oxford\\Oxford, United Kingdom}
}
\def\titlerunning{Correct-by-Construction Control for Stochastic and Uncertain Dynamical Models}
\def\authorrunning{T. Badings, L. Romao, A. Abate, N. Jansen}
\begin{document}
\maketitle

\begin{abstract}
Automated synthesis of correct-by-construction controllers for autonomous systems is crucial for their deployment in safety-critical scenarios. Such autonomous systems are naturally modeled as stochastic dynamical models. The general problem is to compute a controller that provably satisfies a given task, represented as a probabilistic temporal logic specification. However, factors such as stochastic uncertainty, imprecisely known parameters, and hybrid features make this problem challenging. We have developed an abstraction framework that can be used to solve this problem under various modeling assumptions. Our approach is based on a robust finite-state abstraction of the stochastic dynamical model in the form of a Markov decision process with intervals of probabilities (iMDP). We use state-of-the-art verification techniques to compute an optimal policy on the iMDP with guarantees for satisfying the given specification. We then show that, by construction, we can refine this policy into a feedback controller for which these guarantees carry over to the dynamical model. In this short paper, we survey our recent research in this area and highlight two challenges (related to scalability and dealing with nonlinear dynamics) that we aim to address with our ongoing research.
\end{abstract}

\section{Introduction}
\label{sec:Introduction}

Controlled autonomous systems are increasingly deployed in safety-critical settings~\cite{DBLP:journals/tiv/PadenCYYF16}.
When the transitions between states are specified by probabilities, autonomous systems can often be naturally modeled as stochastic dynamical models~\cite{kumar2015stochastic}. 
For deployment in safety-critical settings, controllers for stochastic models must act safely and reliably with respect to desired specifications.
Traditional control design methods use, e.g., Lyapunov functions and optimization to provide guarantees for simple tasks such as stability, convergence, and invariance~\cite{belta2017formal}.
However, alternative methods are needed to give formal guarantees about richer temporal specifications relevant to, for example, safety-critical applications~\cite{DBLP:journals/tac/FanQMNMV22}.

\paragraph{Formal controller synthesis}
Temporal logic is a rich language for specifying the desired behavior of autonomous systems~\cite{platzerLogicsDynamicalSystems2012}.
In particular, probabilistic computation tree logic (PCTL,~\cite{hanssonLogicReasoningTime1994}) is widely used to define temporal requirements on the behavior of probabilistic systems. 
For example, in a motion control problem for an \gls{UAV}, a PCTL formula can specify that, with at least 90\% probability, the \gls{UAV} must safely fly to a target location within 2 minutes without crashing into obstacles (commonly known as a \emph{reach-avoid specification}~\cite{DBLP:conf/hybrid/FisacCTS15}).
Leveraging tools from probabilistic verification~\cite{DBLP:books/daglib/BaierKatoen2008}, the problem is to synthesize a controller that ensures the satisfaction of such a PCTL formula for the model under study~\cite{DBLP:conf/nfm/HahnHZ11}. 
Finite abstractions can make continuous models amenable to techniques and tools from formal verification: by discretizing their state and action spaces, abstractions result in, e.g., finite Markov decision processes (MDPs) that soundly capture the continuous dynamics~\cite{Abate2008probabilisticSystems}.
Verification guarantees on the finite abstraction can thus carry over to the continuous model.

\paragraph{Problem statement}
In this research, we adopt such an abstraction-based approach to controller synthesis for autonomous systems.
Our goal is to compute feedback controllers that \emph{provably satisfy} a given temporal logic specification.
In this short paper, we focus on reach-avoid specifications, but most of our approaches can readily be extended to general PCTL properties~\cite{DBLP:journals/corr/abs-2212-00679}.
We consider the following general problem:
\begin{mdframed}[backgroundcolor=gray!00, nobreak=true, innerleftmargin=8pt, innerrightmargin=8pt]
Given (1) a discrete-time stochastic dynamical model and (2) a reach-avoid specification, compute a \emph{feedback controller} together with a \emph{certificate} in the form of a probability threshold, such that the induced closed-loop system satisfies this specification with at least this certified probability.
\end{mdframed}
In this paper, we survey our recent work in which we have considered this general problem under various modeling assumptions.
First, we provide a general introduction to our abstraction framework.
Thereafter, we summarize our main results from several recent papers~\cite{Badings2023AAAI,Badings2022AAAI,Badings2022JAIR,DBLP:journals/corr/abs-2212-00679}.
Finally, we highlight two key challenges that remain open, and we describe our current research plans that aim to address these changes.

\begin{figure}[t!]
	\centering
	\usetikzlibrary{calc, arrows, arrows.meta, shapes, positioning, shapes.geometric}

\tikzstyle{nodewide} = [rectangle, rounded corners, minimum width=3.6cm, text width=3.3cm, minimum height=1.1cm, text centered, draw=black, fill=plotblue!15]
\tikzstyle{nodewidered} = [rectangle, rounded corners, minimum width=3.6cm, text width=3.3cm, minimum height=1.1cm, text centered, draw=black, ultra thick, fill=red!25]

\small
\resizebox{.6\linewidth}{!}{%
	\begin{tikzpicture}[node distance=6.0cm,->,>=stealth,line width=0.3mm,auto,main node/.style={circle,draw,font=\sffamily\bfseries}]
		
	\newcommand\yshift{-3.8cm}
	
	\node (dyn) [nodewide] 
	{\textbf{Stochastic} \\ \textbf{dynamical model} \\ 
	};
	
	\node (abstract) [nodewide, left of=dyn, xshift=0.5cm] 
	{\textbf{Abstract iMDP}};
	
	\node (guarantees) [nodewide, below of=abstract, yshift=3.8cm] 
	{\textbf{Optimal policy} \\ \textbf{on abstract iMDP}};
	
	\node (policy) [nodewide, below of=dyn, yshift=3.8cm] 
	{\textbf{Certifiably correct} \\ \textbf{feedback controller}};
	
	
	\node (in1) [left of=abstract, xshift=2.0cm] {};
	\node (in2) [left of=guarantees, xshift=2.0cm] {};
	
	\draw [->] (in1) -- (abstract) node [midway, above, align=center, font=\sffamily\small] {State space \\ partition};
	
	\draw [->] (in2) -- (guarantees) node [midway, below, align=center, font=\sffamily\small] {Specification};
	
	\draw [->] (dyn) -- (abstract) node [midway, above, align=center, font=\sffamily\small] {Create \\ abstraction};
	
	
	\draw [->] ($(abstract.south) + (-0.2,0)$) -- ($(guarantees.north) + (-0.2,0)$) node [midway, left, align=center, font=\sffamily\small] {Planning \\ on iMDP};
	
	\draw [->, dashed] ($(guarantees.north) + (0.2,0)$) -- ($(abstract.south) + (0.2,0)$) node [midway, right, align=left, font=\sffamily\small] {$\mathsf{UNSAT?}$ \ \rured{\faIcon{times}} \\ Improve abstraction};
	
	\draw [->] (guarantees) -- (policy) 
	node [midway, above, align=center, font=\sffamily\small] {$\mathsf{SAT?}$ \ \darkgreen{\faIcon{check}}}
	node [midway, below, align=center, font=\sffamily\small] {Extract \\ policy};
	
	\draw [->] ($(dyn.south) + (-0.2,0)$) -- ($(policy.north) + (-0.2,0)$) node [midway, left, align=center, font=\sffamily\small] {State $\state_k$};
	
	\draw [->] ($(policy.north) + (0.2,0)$) -- ($(dyn.south) + (0.2,0)$) node [midway, right, align=center, font=\sffamily\small] {Control \\ input $\control_k$};
 
	\draw [-,decorate,decoration={brace,amplitude=5pt,mirror,raise=2ex}] (-3.7cm, 0.5cm) -- (-7.3cm, 0.5cm)
	node[midway,yshift=3em]{\textbf{Offline planning}};
	\draw [-,decorate,decoration={brace,amplitude=5pt,mirror,raise=2ex}] (1.8cm, 0.5cm) -- (-1.8cm, 0.5cm)
	node[midway,yshift=3em]{\textbf{Online control}};
		
	\end{tikzpicture}
}
	\caption{
		Our overall approach integrated into a safe model-based learning framework.
	}
	\label{fig:Approach}
\end{figure}
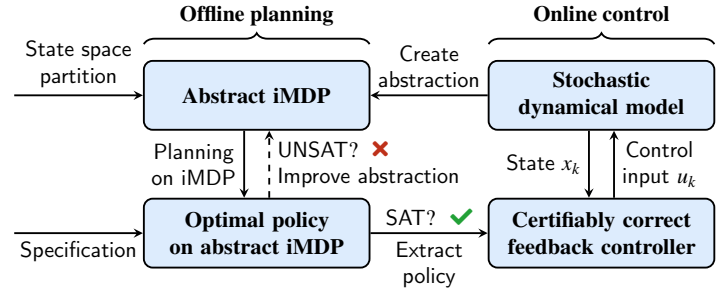

\paragraph{Our abstraction framework}
Our general abstraction framework is shown in \cref{fig:Approach}.
First, we compute a finite-state abstraction of the stochastic dynamical model \cite{DBLP:journals/siamads/SoudjaniA13}, which we obtain from a \emph{partition} of its continuous state space into a set of disjoint convex \emph{regions}.
Actions in this abstraction correspond to continuous control inputs that yield transitions between these regions.
Due to the stochastic noise in the dynamical model, the outcome of an action is stochastic, rendering transitions probabilistic. 
We capture these probabilities in a \gls{MDP}~\cite{DBLP:books/wi/Puterman94}.
A defining characteristic of our approach is that we leverage \emph{backward reachability computations} on the dynamical model to determine which actions are enabled at each discrete region.
By contrast, most other abstraction methods rely on \emph{forward} reachability computations, which are associated with errors that grow with the time horizon of the property (see related work for details).
Our backward scheme avoids such abstraction errors, at the cost of requiring slightly more restrictive assumptions on the model dynamics (see, \eg,~\cite{Badings2022JAIR} for details).

\paragraph{Interval MDPs}
Computing the transition probabilities of the abstraction is subject to estimation errors.
To be \emph{robust} against estimation errors in these probabilities, we use data-driven techniques~\cite{DBLP:journals/arc/CampiCG21,romao2022exact} to compute \emph{upper and lower bounds} on the transition probabilities with a predefined \emph{confidence level}. 
We formalize our abstractions with the \gls{PAC} probability intervals using so-called \glspl{iMDP}, which are an extension of \glspl{MDP} with intervals of probabilities~\cite{DBLP:journals/ai/GivanLD00}. 
Policies for \glspl{iMDP} have to robustly account for \emph{all possible probabilities} within the intervals~\cite{DBLP:conf/cav/PuggelliLSS13, DBLP:conf/cdc/WolffTM12}.
In our implementation, we compute robust policies using robust value iteration within the probabilistic model checker \prism~\cite{DBLP:conf/cav/KwiatkowskaNP11}. 
We show that any policy on the \gls{iMDP} can be refined into a \emph{piecewise linear feedback controller} for the dynamical model.
Crucially, the probability of satisfying the reach-avoid property on the \gls{iMDP} is a \emph{lower bound} on the satisfaction probability for the dynamical model, thus solving the problem above.

\paragraph{Related work}

Abstractions of stochastic models are well-studied~\cite{Abate2008probabilisticSystems,Alur2000discretesystems}, with applications to stochastic hybrid~\cite{DBLP:conf/hybrid/CauchiLLAKC19,LavaeiSAZ21:stochhybrid}, switched~\cite{DBLP:journals/tac/LahijanianAB15}, and partially observable systems~\cite{Badings2021FilterUncertainty,DBLP:conf/adhs/HaesaertNVTAAM18}.
Various tools exist, e.g., \stochy~\cite{DBLP:conf/tacas/CauchiA19} and \probreach~\cite{DBLP:conf/hybrid/ShmarovZ15}.
A distinguishing feature of our abstraction scheme is that we use \emph{backward reachability computations} on the model dynamics to determine the subset of actions enabled in each abstract state.
By contrast, standard abstraction methods typically rely on \emph{forward reachability computations} based on discretizing the control input space.
In particular, such forward methods propagate \emph{sets of states} $\hat{\mathcal{X}} \subset \R^n$ forward through the model dynamics under discretized input $\hat{\bm{u}}_k$ (see the notation from \cref{eq:continuousSystem}).
Since the noise $\noise_k$ is stochastic, this yields a \emph{set of distributions} over successor states, which can be difficult to reason over.
By contrast, with our backward computations, each abstract action yields a \emph{single distribution}, which is independent of where this action was chosen.
However, this requires a higher degree of system controllability, as discussed in more detail in~\cite[Assumption~2]{Badings2022JAIR}.

\section{Correct-by-Construction Control via Formal Abstractions}
\label{sec:Results}

In general, we consider discrete-time, continuous-state dynamical models, where the progression of the state $\state \in \R^n$ depends \emph{linearly} on the current state, on a control input, and on a process noise term.
Given a state $\state_k$ at discrete time $k \in \N$, the successor state $\state_{k+1}$ at time $k+1$ is computed as
\begin{equation}
    \state_{k+1} = A \state_{k} + B \control_{k} + \drift_k + \noise_k,
    \label{eq:continuousSystem}
\end{equation}
with matrices $A$ and $B$, and a continuous control input $\control_k \in \mathcal{U} \subseteq \R^p$ (\ie, action).
The term $\noise_k \in \Delta \subset \R^n$ is an arbitrary additive process noise term, which is an \gls{iid} random variable defined on a probability space $(\Delta, \mathcal{D}, \amsmathbb{P})$, with $\sigma$-algebra $\mathcal{D}$ and probability measure $\amsmathbb{P}$ defined over $\mathcal{D}$. 
A controller (\ie, control policy) $\controller \colon \R^n \times \N \to \cControlSpace$ chooses a control input based on the current state $\state \in \R^n$ and time $k \in \N$.

We now highlight some of the variants of the problem stated in \cref{sec:Introduction} we have considered thus far.

\begin{figure}[t!]
\centering
\begin{minipage}[b]{.48\textwidth} 
    \centering
    \includegraphics[width=\linewidth]
    {Figures/JAIR/UAV_v3.png}
    \caption{UAV reach-avoid problem (goal in green; obstacles in red), plus simulations with the optimal \gls{iMDP}-based controller from initial state $\state_0 = [-14,0,6,0,-6,0]^\top$, under high/low turbulence.}
    \label{fig:UAV}
\end{minipage}
\hfill
\begin{minipage}[b]{.48\textwidth} 
    \centering
    \includegraphics[width=\linewidth]{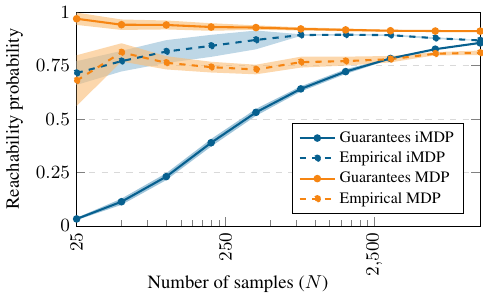}
    \caption{Reach-avoid guarantees on the \glspl{iMDP} (blue) and \glspl{MDP} (orange) for their respective policies, versus the resulting empirical (simulated) performance (dashed lines) on the dynamical system. 
    The empirical performance obtained from the \glspl{MDP} violates the guarantees, whereas that from the \glspl{iMDP} does not.
    }
    \label{fig:UAVrobustness}
\end{minipage}
\end{figure}

\subsection{Stochastic noise of unknown distribution}
\label{subsec:JAIR}
It is commonly assumed that the distribution of the process noise $\noise_k$ is known and/or Gaussian~\cite{DBLP:journals/spm/ParkSQ13}.
However, in many realistic problems, this assumption yields a poor approximation of the uncertainty~\cite{DBLP:journals/trob/BlackmoreOBW10}.
Distributions may even be \textit{unknown}, meaning that one cannot derive a set-bounded or a precise probabilistic representation of the noise. 
In this case, it is generally hard or even impossible to derive \emph{hard guarantees} on the probability that a given controller ensures the satisfaction of a reach-avoid property.

In papers~\cite{Badings2022AAAI,Badings2022JAIR}, we thus consider a variant of the controller synthesis problem from \cref{sec:Introduction} for dynamical systems with additive process noise of an \emph{unknown distribution}.
Specifically, the probability measure $\amsmathbb{P}$ of the noise $\noise_k \in \Delta \subset \R^n$ is unknown but time-invariant.
To deal with this lack of knowledge, we adapt tools from the scenario approach~\cite{DBLP:journals/siamjo/CampiG08,DBLP:journals/arc/CampiCG21} to compute \emph{\gls{PAC}} interval estimates for the transition probabilities of the abstract model based on a finite set of samples of the noise.
We capture these bounds in the transition probability intervals of a so-called interval Markov decision process (iMDP). 
This iMDP is, with a user-specified confidence probability, robust against uncertainty in the transition probabilities, and the tightness of the probability intervals can be controlled through the number of samples. 

In~\cite{Badings2022JAIR}, we use this method to solve a reach-avoid problem for a \gls{UAV} operating under turbulence (we compare scenarios with different turbulence levels), represented by stochastic noise of unknown distribution.
The \gls{UAV} is modeled by a 6D dynamical model (we refer to~\cite{Badings2022JAIR} for the explicit model).
In \cref{fig:UAV}, we show simulations under the optimal controller for two turbulence levels.
Under low noise, the controller prefers the short but narrow path.
On the other hand, under high noise, the longer but safer path is preferred.
Thus, accounting for process noise is important to obtain controllers that are safe.

We also compared our robust \gls{iMDP} approach against a naive \gls{MDP} abstraction.
This \gls{MDP} has the same states and actions as the \gls{iMDP}, but uses precise (frequentist) probabilities.
The maximum reachability probabilities (guarantees) for both methods are shown in \cref{fig:UAVrobustness}.
For every value of $N$, we apply the resulting controllers to the dynamical system in Monte Carlo simulations with $10,000$ iterations to determine the empirical reachability probability.
\cref{fig:UAVrobustness} shows that the non-robust \glspl{MDP} yield \emph{poor and unsafe performance guarantees}: the actual reachability of the controller is much lower than the reachability guarantees obtained from \prism.
By contrast, our robust \gls{iMDP}-based approach consistently yields safe lower bound guarantees on the actual performance of controllers.

\subsection{Set-bounded parameter uncertainty}
\label{subsec:AAAI23}
\begin{figure}[t!]
 \centering
    \includegraphics[width = .6\columnwidth]{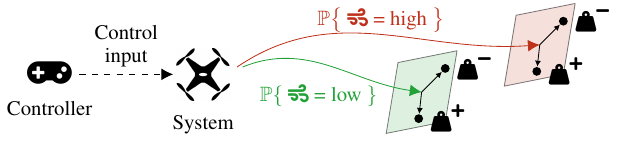}
    \caption{Stochastic uncertainty in the wind (\faIcon{wind}) causes probability distributions over outcomes of controls, while set-bounded uncertainty in the mass (\faIcon{weight-hanging}) of the drone causes state transitions to be nondeterministic.}
	\label{fig:nondeterminism:vs:probabilities}
\end{figure}

The approach described in \cref{subsec:JAIR} requires \emph{precise knowledge} of the model parameters (namely, the matrices $A$ and $B$).
However, in many realistic cases, there is \emph{epistemic uncertainty} about the precise values of these parameters.
For example, consider again the \gls{UAV} from \cref{subsec:JAIR}.
As shown in \cref{fig:nondeterminism:vs:probabilities}, the drone's dynamics depend on uncertain factors, such as the wind and the drone's mass.
We assumed that the wind is adequately described by a probabilistic model, reflected in the process noise $\noise_k$.
Now, let us assume we know that the drone's mass lies between $0.75$--$\SI{1.25}{\kilogram}$, but we do not have information about the likelihood of each value, so employing a probabilistic model is unrealistic. 
Thus, we treat epistemic uncertainty in such imprecisely known parameters (in this case, the mass) using a \emph{nondeterministic framework} instead.

We have recently extended our abstraction framework in~\cite{Badings2023AAAI} to capture both stochastic noise and set-bounded uncertain parameters.
Specifically, we synthesize a controller that (1) is \emph{robust against nondeterminism} due to parameter uncertainty and (2) \emph{reasons over probabilities} derived from stochastic noise.
In other words, the controller must satisfy a given specification \emph{under any possible outcome of the nondeterminism} (robustness) and \emph{with at least a certain probability regarding the stochastic noise} (reasoning over probabilities).
As before, we wish to synthesize a controller with a \emph{\gls{PAC}}-style guarantee: we wish to find a controller that satisfies a reach-avoid specification with at least a desired lower bound threshold probability, and (because our algorithm involves random sampling) that claim should hold with at least a predefined confidence level.

Our experiments in~\cite{Badings2023AAAI} show that we can synthesize controllers that are robust against uncertainty and, in particular, against deviations in the model parameters.
Moreover, we show that our method can be used to faithfully capture any uncertainty or error term in the dynamical model that is represented by a bounded set, thus opening the door for the abstraction of nonlinear systems.

\subsection{Markov jump linear systems}
\label{subsec:QEST23}

Our approaches described so far are limited to systems with purely continuous dynamics.
Thus, these approaches are incompatible with cyber-physical systems, which are characterized by the coupling of digital (discrete) with physical (continuous) components. 
This results in a \emph{hybrid system} that can jump between discrete modes of operation, each of which is characterized by its own continuous dynamics~\cite{LavaeiSAZ21:stochhybrid}.

To alleviate this restriction, we have extended our abstraction framework in \cite{DBLP:journals/corr/abs-2212-00679} to Markov jump linear systems (MJLSs), which are a well-known class of stochastic, hybrid models suitable for capturing the behavior of cyber-physical systems~\cite{docostaDiscreteTimeMarkovJump2005}. 
An MJLS consists of a finite set of linear dynamics defined by \cref{eq:continuousSystem} (also called \emph{operational modes}), where jumps between these modes are governed by a \gls{MC}.
If mode jumping can be controlled, the jumps are governed by an \gls{MDP}.
Due to the jumping between modes, the overall dynamics of an MJLS are nonlinear, making controller synthesis challenging.
For brevity, we refer to~\cite{DBLP:journals/corr/abs-2212-00679} for further results in this problem setting.

\section{Current research directions}
As discussed above, we have considered the general problem in \cref{sec:Introduction} under various model assumptions.
At the same time, each of those settings suffers from its limitations and necessary assumptions, so the general problem of \emph{optimal control under uncertainty} is far from solved.
In this section, we discuss two key limitations of our current framework, which are related to scalability and to linearity of the dynamics.
Moreover, we describe how we try to address both of these challenges with our ongoing research.

\subsection{Neural-guided abstraction of nonlinear systems}
\label{subsec:Neural}

Thus far, our research has focused on dynamical models with linear dynamics.
Extensions to nonlinear dynamical systems are non-trivial and may require more involved reachability computations~\cite{DBLP:conf/cdc/BansalCHT17,DBLP:conf/cav/ChenAS13}.
Specifically, the challenge is that the backward reachability computations involved in our approach may become non-convex under nonlinear dynamics.

\paragraph{Neural network partitioning}
A recent paper~\cite{DBLP:conf/nips/AbateEG22} has proposed to use feedforward neural networks to \emph{learn} state space partitions for nonlinear dynamical models into polyhedral regions.
Inspired by this approach, we are developing an abstraction procedure for nonlinear stochastic dynamical models, which (1) learns a polyhedral state space partition using a neural network, and (2) constructs a piecewise linear approximation of the nonlinear dynamics based on this partition.
By defining the loss function for the neural network such that it minimizes the linearization error across the partition, we hope to find smarter partitions (into fewer elements and of better geometry) than the rectangular ones we employed thus far.

\paragraph{Abstraction of linearized dynamics}
To account for the error caused by the linearization, we add a set-bounded nondeterministic disturbance to the linearized dynamics.
As we have shown in~\cite{Badings2023AAAI}, we can robustly capture this set-bounded disturbance in an \gls{iMDP} abstraction.
However, the quality of the abstraction largely depends on the size of the disturbance representing the linearization error.
Thus, the main challenge with this approach is to obtain a tight, set-bounded representation of the linearization error.

\subsection{Abstractions of polyhedral Lyapunov functions}
\label{subsec:Lyapunov}

Discrete abstractions are computationally expensive in general due to the discretization of the state space. 
For example, the number of abstract states scales exponentially with the dimension of the state space, commonly called the \emph{curse of dimensionality}.
Moreover, adding robustness to multiple sources of uncertainty (as we have done in~\cite{Badings2023AAAI} further increases the number of transitions modeled in the abstract model).
Thus, finding ways to reduce the complexity of abstraction while keeping their expressivity is a challenging direction for further research.

\paragraph{Abstraction of Lyapunov functions}
Inspired by~\cite{DBLP:conf/adhs/DingLB12} and the large body of literature on Lyapunov and Barrier functions~\cite{DBLP:conf/eucc/AmesCENST19}, we are developing a method for abstracting stochastic dynamical systems using Lyapunov functions.
Specifically, we wish to generate an abstract model whose states represent annuli of the sublevel sets of a Lyapunov function.
A similar approach was used by~\cite{DBLP:conf/adhs/DingLB12}.
However, the approach by~\cite{DBLP:conf/adhs/DingLB12} relies on a \emph{strict decrease condition} on the Lyapunov function and is therefore restricted to \emph{nonstochastic} linear systems only.
Instead, we believe that our abstraction procedure based on backward reachability analysis can be used to construct \emph{sound} abstractions of sublevel sets of Lyapunov functions.

\paragraph{Complexity is independent of state dimension}
This envisioned abstraction of Lyapunov sublevel sets avoids the need for an exhaustive partitioning of the state space.
Notably, the number of states in the envisioned abstraction is \emph{independent of the dimension of the state space}.
Thus, we believe that this approach may significantly reduce the computational complexity of the abstraction.

\section{Conclusions and Future Work}
In this short paper, we have surveyed our recent research on abstraction-based controller synthesis for stochastic and uncertain dynamical models.
Based on a robust finite-state abstraction in the form of an \gls{iMDP}, we are able to synthesize controllers for dynamical models that \emph{provably satisfy} given temporal logic specifications, such as reach-avoid tasks.
We have considered this general problem under various modeling assumptions, including unknown noise distributions, imprecisely known model parameters, and hybrid features.
Moreover, we have highlighted two key challenges that are related to scalability and extensions to nonlinear systems.
With our ongoing research, we aim to address these challenges.

\paragraph{Acknowledgements} 
This research has been partially funded by NWO grant NWA.1160.18.238 (PrimaVera), by EPSRC IAA Award EP/X525777/1, and by the ERC Starting Grant 101077178 (DEUCE).

\bibliographystyle{eptcs}
\bibliography{generic}
\end{document}